\documentclass[12pt]{iopart}
\usepackage{iopams}
\begin{document}
\title[Quasiperiodic packings of $G$-clusters and Baake-Moody sets]{Quasiperiodic 
packings of $G$-clusters and Baake-Moody 
sets}
\author{Nicolae Cotfas}
\address{Faculty of Physics, University of Bucharest,
PO Box 76-54, Postal Office 76, Bucharest, Romania,\\ 
E-mail address: ncotfas@yahoo.com\ \ \ 
'http://fpcm5.fizica.unibuc.ro/\~{ }ncotfas}
\begin{abstract}
The diffraction pattern of a quasicrystal admits as
symmetry group a finite group $G$, and there exists a
$G$-cluster ${\mathcal C}$ (a union of orbits of $G$) 
such that the quasicrystal can be regarded as a 
quasiperiodic packing of copies of ${\mathcal C}$,
generally, partially occupied. 
On the other hand, by starting from the
$G$-cluster ${\mathcal C}$ we can define in a canonical way
a permutation representation of $G$ in a higher 
dimensional space, decompose this space into the orthogonal
sum of two $G$-invariant subspaces and use the strip 
projection method in order to define a pattern which
can also be regarded as a quasiperiodic 
packing of copies of ${\mathcal C}$, generally, partially occupied.
This mathematical algorithm is useful in quasicrystal physics,
but the dimension of the superspace we have to use
in the case of a two or three-shell cluster is rather large. 
We show that the generalization concerning the strip projection
method proposed by Baake and Moody [{\it Proc. 
Int. Conf. Aperiodic' 97} (Alpe d'Huez, 27-31 August, 1997)
ed M de Boissieu {\it et al.} (Singapore: World Scientific, 1999) 
pp 9-20] allows to reduce this dimension, and present some examples.
\end{abstract}
\maketitle
\section{Introduction}

The diffraction pattern of a quasicrystal contains a discrete set of
intense Bragg peaks invariant under a finite group $G,$ and the
high-resolution electron microscopy suggests that the quasicrystal can be
regarded as a packing of (partially occupied) copies of a well-defined 
$G$-invariant atomic cluster ${\mathcal C}$. From a mathematical point of
view, the cluster ${\mathcal C}$ can be defined as a finite union of orbits 
of $G$, and there exists an algorithm \cite{C1,C2} which leads from ${\mathcal C}$ 
directly to a pattern ${\mathcal Q}$ which can be regarded as a union of
interpenetrating partially occupied translations of ${\mathcal C}$ 
(the neighbours of each point $x\in {\mathcal Q}$ belong to the set 
$x+{\mathcal C}=\{ x+y\ |\ y\in {\mathcal C}\}$). This algorithm, based on
the strip projection method and group theory, represents an extended
version of the model proposed by Katz \& Duneau and independently by Elser 
for the icosahedral quasicrystals.

Unfortunately, the dimension of the superspace used in the case of a two or
three-shell icosahedral cluster is rather large. The main purpose of this
article is to present a way to reduce this dimension. 
It is based on the 
extension of the notion of model set proposed by Baake and Moody \cite{B}.
This extension increases the power of the strip projection method, 
and allows to define a larger class of quasiperiodic patterns,
very useful in quasicrystal mathematics. 
We call them {\em Baake-Moody sets}. 
Some examples are presented in order to illustrate the
theoretic considerations.

\section{Baake-Moody sets}

Let ${\mathbb{E}}_k=(\mathbb{R}^k,\langle , \rangle )$ be the usual 
$k$-dimensional Euclidean space,
$E$ and $E^\perp $ be two orthogonal
subspaces such that $\mathbb{E}_k=E\oplus E^\perp $, and let 
\begin{equation}
\mathbb{L}=\kappa \mathbb{Z}^k\qquad  
\mathbb{K}=[0,\kappa ]^k=\{ (x_1,x_2,...,x_k)\ |\
0\leq x_i\leq \kappa \ {\rm for\ all\ } i\}
\end{equation}
where $\kappa \in (0,\infty )$ is a fixed constant.
For each $x\in \mathbb{E}_k$ there exist $x^\parallel \in E$  and 
$x^\perp \in E^\perp $ uniquely determined such that 
$x=x^\parallel +x^\perp .$ The mappings
\begin{equation} \fl
\pi :\mathbb{E}_k\longrightarrow \mathbb{E}_k:x\mapsto \pi x=x^\parallel \qquad 
\pi ^\perp :\mathbb{E}_k\longrightarrow \mathbb{E}_k:x\mapsto \pi ^\perp x=x^\perp 
\end{equation}
are the corresponding orthogonal projectors.

By using the bounded set $K=\pi ^\perp (\mathbb{K})$  we define in terms of the
{\em strip projection method} \cite{K} the discrete set
\begin{equation}
{\mathcal Q} =\left\{ \pi x\ \left| \ x\in \mathbb{L} , \ \pi ^\perp x\in K
\right.\right\}
\end{equation}
formed by the projection on $E$ of all the points of $\mathbb{L}$ lying 
in the {\em strip} $K+E=\{ x+y\ |\ x\in K, \ y\in E \}.$ 

\begin{figure}
\setlength{\unitlength}{1mm}
\begin{picture}(70,50)(-25,0)
\put(62,19){$E$}
\put(0,19){$E'$}
\put(9,42){$E''$}
\put(24.5,43){$E^\perp $}
\put(18.2,37){$K$}
\put(7,22){$K_0$}
\put(8,32){$K_i$}
\put(8.5,14.3){$K_j$}
\put(57,26){${\mathcal E}$}
\put(38,32){${\mathcal E}_i=z_i+{\mathcal E}$}
\put(38,14){${\mathcal E}_j=z_j+{\mathcal E}$}
\put(4,0){\line(0,1){38}}
\put(4,0){\line(3,1){20}}
\put(4,38){\line(3,1){20}}
\put(24,6.8){\line(0,1){38}}
\put(4,19){\line(1,0){40}}
\put(24,25.5){\line(1,0){40}}
\put(44,19){\line(3,1){20}}
\put(4,28){\line(1,0){40}}
\put(4,28){\line(3,1){20}}
\put(24,34.5){\line(1,0){40}}
\put(44,28){\line(3,1){20}}
\put(4,10){\line(1,0){40}}
\put(4,10){\line(3,1){20}}
\put(24,16.5){\line(1,0){40}}
\put(44,10){\line(3,1){20}}
\put(9,5){\line(4,1){10}}
\put(9,5){\line(-1,4){4.3}}
\put(4.8,22.4){\line(1,4){3.5}}
\put(18.8,7.3){\line(1,4){4}}
\put(22.5,23.4){\line(-1,3){5}}
\put(8.7,36.4){\line(4,1){8.5}}
\linethickness{0.4mm}
\put(14,22.3){\line(1,0){50}}
\put(0,17.5){\line(3,1){24.1}}
\put(0,17.4){\line(3,1){24.1}}
\put(0,17.6){\line(3,1){24.1}}
\put(5.7,19.2){\line(3,1){16.4}}
\put(5.7,19.7){\line(3,1){16.4}}
\put(6.5,29){\line(3,1){12.8}}
\put(6.5,29.1){\line(3,1){12.8}}
\put(6.5,29.2){\line(3,1){12.8}}
\put(6.5,28.9){\line(3,1){12.8}}
\put(6.5,28.8){\line(3,1){12.8}}
\put(7.7,11.4){\line(3,1){13}}
\put(7.7,11.5){\line(3,1){13}}
\put(7.7,11.6){\line(3,1){13}}
\put(7.7,11.2){\line(3,1){13}}
\put(7.7,11.3){\line(3,1){13}}
\linethickness{0.4mm}
\put(14,3.5){\line(0,1){42}}
\end{picture}
\caption{The decompositions $\mathbb{E}_k=E\oplus E^\perp=
E\oplus E'\oplus E''={\mathcal E}\oplus E''$.}
\end{figure}
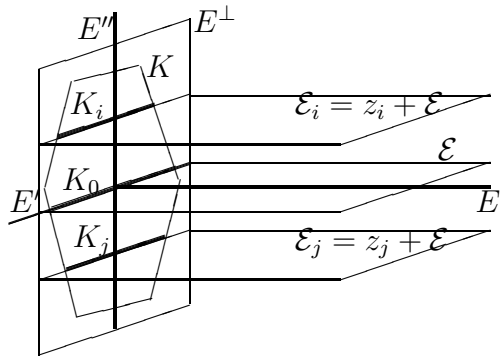

It is known \cite{S} that any $\mathbb{Z}$-module $M\subset \mathbb{R}^l$ is the direct 
sum of a lattice $M_d$ of rank $d$ and a $\mathbb{Z}$-module $M_s$ dense in a vector 
subspace of dimension $s$, where $d+s$ is the dimension of the subspace  
generated by $L$ in $\mathbb{R}^l.$ In view of this result  
the $\mathbb{Z}$-module $\mathbb{L}^\perp =\pi ^\perp (\mathbb{L})$ is the direct sum 
$\mathbb{L}^\perp ={\mathcal L}'\oplus {\mathcal D}$ of a lattice ${\mathcal D}$
of rank $d$ and a $\mathbb{Z}$-module ${\mathcal L}'$ dense in a subspace
$E'\subset E^\perp $ of dimension $s$, where $d+s={\rm dim}\, E^\perp $.
In this decomposition the space $E'$ is uniquely determined and we denote
by $E''$ its orthogonal complement in $E^\perp $
\[ E''=\left\{ \left.  x\in E^\perp \  \right| \ \langle x,y\rangle =0\ 
{\rm for \ any}\ y\in E' \right\} . \]
We get $\mathbb{E}_k=E\oplus E'\oplus E''$.
For each $x\in \mathbb{E}_k$ there exist $x^\parallel \in E$, $x'\in E'$ and 
$x''\in E''$ uniquely determined such that $x=x^\parallel +x'+x''$.
The mappings
$\pi ':\mathbb{E}_k\longrightarrow \mathbb{E}_k, \ \pi ' x=x' $ and 
$\pi '':\mathbb{E}_k\longrightarrow \mathbb{E}_k,\ \pi '' x=x''$
are the orthogonal projectors corresponding to $E'$ and $E''$.

One can prove \cite{C3} that the projection ${\mathcal L}=(\pi +\pi ')(\mathbb{L})$ 
of the lattice $\mathbb{L}$ on the space ${\mathcal E}=E\oplus E'$ is a lattice 
in ${\mathcal E}$, $\pi $ restricted to ${\mathcal L}$ is injective and
$\pi ' ({\mathcal L})$ is dense in $E'.$ It follows that 
the collection of spaces and mappings 
\begin{equation}\begin{array}{ccccccc}
\pi x \leftarrow x 
&:E& \stackrel{\pi }\longleftarrow & {\mathcal E}
& \stackrel{\pi ' }\longrightarrow & E' :& x \rightarrow \pi ' x\\
&&&\cup &&& \\
&&& {\mathcal L} &&&
\end{array}
\end{equation} 
is a {\it cut and project scheme} \cite{B,M}.      

The lattice $L=\mathbb{L}\cap {\mathcal E}$ is a sublattice of ${\mathcal L}$,
and necessarily $[{\mathcal L}:L]$ is finite. 
The projection $\mathbb{L}''=\pi ''(\mathbb{L})$ of $\mathbb{L}$ on $E''$ is a discrete countable
set. Let ${\mathcal Z}=\{ z_i\ |\ i\in \mathbb{Z} \}$ be a subset of ${\mathcal L}$ such
that $\mathbb{L}''=\pi ''({\mathcal Z})$ and $\pi ''u_i\not=\pi ''u_j$ for $i\not=j$. 
The lattice $\mathbb{L}$ is contained in the union of the cosets
${\mathcal E}_i=z_i+{\mathcal E}=\{ z_i+x\ |\  x\in {\mathcal E}\}$ 
\begin{equation}
\mathbb{L}\subset \bigcup_{i\in\mathbb{Z}} {\mathcal E}_i.
\end{equation}
Since $\mathbb{L}\cap {\mathcal E}_i=z_i+L$ the set
\begin{equation}
{\mathcal L}_i=(\pi +\pi ')(\mathbb{L}\cap {\mathcal E}_i)=
(\pi +\pi ')z_i+L
\end{equation}
is a coset of $L$ in ${\mathcal L}$ for any $i\in \mathbb{Z}$.

Only for a finite number of cosets ${\mathcal E}_i$ the intersection 
\begin{equation}
K_i=K\cap {\mathcal E}_i=\pi ^\perp (\mathbb{K}\cap {\mathcal E}_i)
\subset \pi ''z_i+E'
\end{equation}
is non-empty .
By changing the indexation of the elements of ${\mathcal Z}$
if necessary, we can assume that the subset of $E'$
\begin{equation}
{\mathcal K}_i=\pi '(K_i)=\pi '(\mathbb{K}\cap {\mathcal E}_i)\subset E'
\end{equation}
has a non-empty interior only for $i\in \{ 1,...,m\}.$ 
The `polyhedral' set ${\mathcal K}_i$ satisfies the conditions:
\begin{itemize}
\item[(a)] ${\mathcal K}_i\subset E'$ is compact;
\item[(b)] ${\mathcal K}_i=\overline{\rm int({\mathcal K}_i)}$;
\item[(c)] The boundary of ${\mathcal K}_i$ has Lebesgue measure $0$
\end{itemize}
for any $i\in \{ 1,...,m\}.$
This allows us to use the Baake-Moody generalization 
of the notion of model set presented in \cite{B} in order to define the set
\begin{equation}
\Lambda =\bigcup_{i=1}^m
\left\{ \pi x\ \left| \ x\in {\mathcal L}_i , \ \pi ' x\in {\mathcal K}_i
\right.\right\}
\end{equation}
which we call a {\em Baake-Moody set}. 

One can remark that $\Lambda ={\mathcal Q}$, 
and hence we have re-defined our pattern ${\mathcal Q}$
as a Baake-Moody set by using the superspace ${\mathcal E}$ of dimension,
generally, smaller than the dimension $k$ of the initial superspace $\mathbb{E}_k$.
The main difficulty in this new approach is the determination of the 
`atomic surfaces' ${\mathcal K}_i$.

\section{Packings of $G$-clusters obtained by projection}

In this section we review our method to obtain packings of clusters by 
projection. It is a direct generalization of the model proposed by 
Katz \& Duneau \cite{K} and independently by Elser \cite{E} 
for icosahedral quasicrystals.

Let $\{ g:\mathbb{E}_n\longrightarrow \mathbb{E}_n\ |\ \ g\in G\ \}$ be an
orthogonal $\mathbb{R}$-irreducible faithful representation of a finite 
group $G$ in the `physical' space $\mathbb{E}_n$ 
and let $S\subset \mathbb{E}_n$ be a finite non-empty set 
which does not contain the null vector. Any finite union of orbits
of $G$ is called a {\it $G$-cluster.} Particularly,
\begin{equation}\begin{array}{l}
{\mathcal C}=\bigcup_{r\in S}Gr\, \cup \, \bigcup_{r\in S}G(-r)=
\{ e_1,e_2,...,e_k,-e_1,-e_2,...,-e_k\}
\end{array}\end{equation}
where $Gr=\{ gr\ |\ g\in G\}$, 
is the $G$-cluster symmetric with respect to the origin generated 
by $S.$ 

Let $e_i=(e_{i1},e_{i2},...,e_{in})$, and let
$\varepsilon _1=(1,0,...,0),\ \varepsilon _2=(0,1,0,...,0),\ ...,\
\varepsilon _k=(0,...,0,1)$ be the canonical basis of $\mathbb{E}_k$.
For each $g\in G,$ there exist
the numbers $s_1^g,\ s_2^g,...,s_k^g\in \{ -1;\ 1\}$ and a permutation of 
the set $\{ 1,2,...,k\}$ denoted also by $g$ such that,
\begin{equation}\label{sp}
ge_j=s_{g(j)}^ge_{g(j)}\qquad {\rm for\ all\ }
j\in \{ 1,2,...,k\}. 
\end{equation}
{\bf Theorem 1.} \cite{C1,C2} {\it  
The group $G$ can be identified with the group of permutations
$ \left\{ \ {\mathcal C}\longrightarrow {\mathcal C}:r\mapsto gr\ 
\rule[-2pt]{1pt}{12pt}\ g\in G\ \right\}$
and the formula 
\begin{equation}
g\varepsilon _j=s_{g(j)}^g\varepsilon _{g(j)}\qquad { for\ all\ }
j\in \{ 1,2,...,k\}. 
\end{equation}
defines the orthogonal representation
\begin{equation}
 g(x_1,x_2,...,x_k)=(s_1^gx_{g^{-1}(1)},s_2^gx_{g^{-1}(2)},...,
s_k^gx_{g^{-1}(k)})
\end{equation}
of $G$ in $\mathbb{E}_k.$}\\[3mm]
{\bf Theorem 2.} \cite{C1,C2} {\it The subspaces
\begin{equation}\begin{array}{l}
E=\left\{ \ (<r,e_1>,<r,e_2>,...,<r,e_k>)\ |\ \  
                                       r\in \mathbb{E}_n\ \right\}\\[2mm]
E^{\perp }=\left\{ \ (x_1,x_2,...,x_k)\in \mathbb{E}_k\ \left| \  
\sum_{i=1}^kx_i\, e_i=0\right. \right\}
\end{array} \end{equation}
of $\mathbb{E}_k$ are $G$-invariant, orthogonal, and
$\mathbb{E}_k=E\oplus E^{\perp }.$ }\\[3mm]
{\bf Theorem 3.} \cite{C1,C2} {\it The vectors 
$v_1=\varrho (e_{11},e_{21},...,e_{k1})$,...,
$v_n=\varrho (e_{1n},e_{2n},...,e_{kn})$, where
$\varrho =1/\sqrt{(e_{11})^2+(e_{21})^2+...+(e_{k1})^2}$
form an orthonormal basis of $E$.}\\[3mm]
{\bf Theorem 4.} \cite{C1,C2} {\it The subduced representation of $G$ in $E$
is equivalent with the representation of $G$ in $\mathbb{E}_n,$ and the isomorphism
\begin{equation}
\varphi :\mathbb{E}_n\longrightarrow E\qquad 
\varphi (r)=(\varrho <r,e_1>,\varrho <r,e_2>,...,\varrho <r,e_k>)
\end{equation}
with the property 
$\varphi (\alpha _1,\alpha _2,...,\alpha _n)
=\alpha _1v_1+\alpha _2v_2+...+\alpha _nv_n$ allows us to identify 
the `physical' space $\mathbb{E}_n$ with the subspace $E$ of $\mathbb{E}_k$.}\\[3mm]
{\bf Theorem 5.} \cite{C1,C2} {\it The matrix of the orthogonal projector
 \ $\pi :\mathbb{E}_k\longrightarrow \mathbb{E}_k$ corresponding to $E$ in the basis
$\{ \varepsilon _1,\varepsilon _2,...,\varepsilon _k\}$  is}
\begin{equation}
 \pi =\varrho ^2\left( \begin{array}{llll}
\langle e_1,e_1\rangle & \langle e_1,e_2\rangle & ... & \langle e_1,e_k\rangle \\
\langle e_2,e_1\rangle & \langle e_2,e_2\rangle & ... & \langle e_2,e_k\rangle \\
...& ... & ... & ...\\
\langle e_k,e_1\rangle & \langle e_k,e_2\rangle & ... & \langle e_k,e_k\rangle 
\end{array}  \right) .
\end{equation}

Let $\kappa =1/\varrho $, \ $\mathbb{L}=\kappa \mathbb{Z}^k$, \ 
$\mathbb{K}=[0,\kappa ]^k$, and let $K=\pi ^\perp (\mathbb{K})$, where
$\pi ^\perp :\mathbb{E}_k\longrightarrow \mathbb{E}_k$, 
$\pi ^\perp x=x-\pi x$ 
is the orthogonal projector corresponding to $E^\perp $.\\[3mm]
{\bf Theorem 6.} \cite{C1,C2} {\it The $\mathbb{Z}$-module $\mathbb{L}\subset \mathbb{E}_k$
is $G$-invariant, \ $\pi (\kappa \varepsilon _i)=\varphi (e_i)$,
that is, $\pi (\kappa \varepsilon _i)=e_i$ if we take into consideration 
the identification} $\varphi :\mathbb{E}_n\longrightarrow E,$ and
\begin{equation}
\pi (\mathbb{L})=\mathbb{Z}e_1+\mathbb{Z}e_2+...+\mathbb{Z}e_k.
\end{equation}

The pattern defined by using the strip projection method \cite{K}
\begin{equation}
{\mathcal Q}=\left\{ \left. \pi  x\ \right| \ 
x\in \mathbb{L},\ \pi ^\perp x\in K\right\}
\end{equation}
can be regarded as a union of interpenetrating 
partially occupied copies of ${\mathcal C}$. 
For each point $\pi x\in {\mathcal Q}$ the set of all the arithmetic neighbours 
of $\pi x$
\[ \{ \pi y\ |\  y\in \{ x+\kappa \varepsilon _1,...,x+\kappa \varepsilon _k,
x-\kappa \varepsilon _1,...,x-\kappa \varepsilon _k\},  \pi ^\perp y\in K\}\]
is contained in the translated copy 
\[ \{ \pi x +e_1,...,\pi x+e_k,\pi x -e_1,...,\pi x-e_k\}=\pi x +{\mathcal C} \]
of the $G$-cluster ${\mathcal C}$.
A fragment of ${\mathcal Q}$ can be 
obtained by using, for example, the algorithm presented in \cite{V}. 

The method presented in the previous section allows to 
re-define ${\mathcal Q}$ as a Baake-Moody set by using, generally, a smaller dimensional
superspace. Some exemples are presented in sections 4-7.

\section{A 2D Penrose pattern}

The relations 
\begin{equation}\label{rep1} 
a(x,y)=(cx-sy,sx+cy)\qquad \qquad  b(x,y)=(x,-y) 
\end{equation} 
where  
$c=\cos (\pi /5)=(1+\sqrt{5})/4,$  
$s=\sin (\pi /5)=\sqrt{10-2\sqrt{5}}/4$ 
define the usual two-dimensional representation of the dihedral group  
\[ D_{10}=\left< a,\; b\ \left|\ a^{10}=b^2=(ab)^2=e\right. \right>.\] 
Let $\varepsilon _1=(1,0,...,0),\ \varepsilon _2=(0,1,0,...,0),\ ...,\
\varepsilon _5=(0,...,0,1)$ be the canonical basis of $\mathbb{E}_5$, and let
$c'=\cos (2\pi /5)=(\sqrt{5}-1)/4,$ 
$s'=\sin (2\pi /5)=\sqrt{10+2\sqrt{5}}/4.$ 
The $D_{10}$-cluster (containing only one orbit)
generated by the set $S=\{ (1,0)\}$ is
\[ {\mathcal C}=D_{10}(1,0)=\{ e_1,e_2,e_3,e_4,e_5,-e_1,-e_2,-e_3,-e_4,-e_5\} \] 
where 
$e_1=(1,0),$ $e_2=(c',s'),$ $e_3=(-c,s),$ $e_4=(-c,-s),$ 
$e_5=(c',-s').$ It is formed by the vertices of a regular decagon.

The action of $a$ and $b$ on 
${\mathcal C}$ is described by the signed permutations
\begin{equation}
\fl  a=\left( \begin{array}{rrrrr}
                e_1 & e_2 & e_3 & e_4 & e_5\\
                -e_4 & -e_5 & -e_1 & -e_2 & -e_3 
                \end{array} \right)
\qquad b=\left( \begin{array}{rrrrr}
                e_1 & e_2 & e_3 & e_4 & e_5\\
                e_1 & e_5 & e_4 & e_3 & e_2  
                \end{array} \right) 
\end{equation}
and the corresponding transformations $a,\ b:\mathbb{E}_6\longrightarrow \mathbb{E}_6$
\begin{equation}
\fl a=\left( \begin{array}{rrrrr}
                \varepsilon _1 & \varepsilon _2 & \varepsilon _3 & 
                \varepsilon _4 & \varepsilon _5 \\
                \varepsilon _2 & \varepsilon _3 & \varepsilon _4 & 
                \varepsilon _5 & -\varepsilon _1
                \end{array} \right)
\qquad  b=\left( \begin{array}{rrrrr}
     \varepsilon _1 & \varepsilon _2 & \varepsilon _3 & \varepsilon _4 & 
     \varepsilon _5\\
     \varepsilon _1 & -\varepsilon _5 & -\varepsilon _4 & -\varepsilon _3 & 
     -\varepsilon _2
                \end{array} \right) 
\end{equation}
generate the orthogonal representation of $D_{10}$ in $\mathbb{E}_5$ 
\begin{equation} \begin{array}{l} 
a(x_1,x_2,x_3,x_4,x_5)=(-x_3,-x_4,-x_5,-x_1,-x_2)\\ 
b(x_1,x_2,x_3,x_4,x_5)=(x_1,x_5,x_4,x_3,x_2). 
\end{array}  
\end{equation} 
 
The vectors 
$v_1=\varrho (1,c',-c,-c,c'),$ $v_2=\varrho (0,s',s,-s,-s'),$ 
where $\varrho =\sqrt{2/5},$ form an orthonormal basis of the  
$D_{10}$-invariant subspace 
\begin{equation} 
 E= 
\left\{ (<r,e_1>,<r,e_2>,...,<r,e_5>)\ 
|\ \ r\in \mathbb{E}_2\ \right\} 
\end{equation} 
and the isometry (which is an isomorphism of representations) 
\begin{equation} 
\varphi :\mathbb{E}_2\longrightarrow E: r\mapsto 
(\varrho <r,e_1>,\varrho <r,e_2>,...,\varrho <r,e_5>) 
\end{equation} 
with the property $\varphi (\alpha ,\beta )=\alpha v_1+\beta v_2$ allows us  
to identify the physical space $\mathbb{E}_2$ with the subspace $E$ of $\mathbb{E}_5$. 
The matrices of the orthogonal projectors 
$\pi ,\, \pi ^\perp :\mathbb{E}_5\longrightarrow \mathbb{E}_5$
corresponding to $E$ and 
\begin{equation}
E^\perp =\{ x\in \mathbb{E}_5\ |\ \langle x,y\rangle =0 \
{\rm for\ all\ } y\in E\}
\end{equation}
in the basis $\{ \varepsilon _1,...,\varepsilon _5\}$ are
\begin{equation} 
\pi ={\mathcal M}(2/5,-\tau '/5,-\tau /5)\qquad 
\pi ^{\perp }={\mathcal M}(3/5,\tau '/5,\tau /5)
\end{equation} 
where  $\tau =(1+\sqrt{5})/2$, \ $\tau ' =(1-\sqrt{5})/2$ and
\begin{equation} 
{\mathcal M}(\alpha ,\beta ,\gamma )=\left( \begin{array}{rrrrr} 
\alpha &\beta &\gamma &\gamma &\beta  \\ 
\beta & \alpha &\beta &\gamma &\gamma \\  
\gamma &\beta & \alpha &\beta &\gamma \\ 
\gamma &\gamma &\beta &\alpha &\beta  \\ 
\beta &\gamma &\gamma &\beta &\alpha  
	     \end{array} \right) . 
\end{equation} 

Let $\kappa =1/\varrho =\sqrt{5/2}$, $\mathbb{L}=\kappa \mathbb{Z}^5$, $\mathbb{K}=[0,\kappa ]^5$,
and let $K=\pi ^\perp (\mathbb{K})$. 
The pattern defined in terms of the strip projection method
\begin{equation}
{\mathcal Q}=\left\{ \left. \pi  x\ \right| \ 
x\in \mathbb{L},\ \pi ^\perp x\in K\right\} 
\end{equation}
is the set of vertices of a {\em 2D Penrose (singular) pattern} \cite{K}.
For each $i\in \{ 1,2,3,4,5\}$ we have $\pi (\kappa \varepsilon _i)
=\varphi (e_i)$, that is, $\pi (\kappa \varepsilon _i)=e_i$ if we
use the identification of $\mathbb{E}_2$ with $E$. Since the arithmetic
neighbours of each point $\pi x\in {\mathcal Q}$ 
belong to the set $\pi x+{\mathcal C}$
we can regard ${\mathcal Q}$ as a quasiperiodic packing of partially 
occupied copies of the $D_{10}$-cluster ${\mathcal C}$, that is, a 
quasiperiodic packing of decagons.

We can re-define ${\mathcal Q}$ as a Baake-Moody set by using the
general method presented in section 2. In this case, we have  to
use the decomposition $E^\perp =E'\oplus E''$ corresponding to the 
orthogonal projectors 
\[ \pi '={\mathcal M}( 2/5,-\tau /5,-\tau '/5)\qquad 
\pi ''={\mathcal M}(1/5,1/5,1/5).\]
We get 
\begin{equation} \fl
{\mathcal E}=E\oplus E'=\{ (x_1,x_2,x_3,x_4,x_5)\in \mathbb{E}_5\ |\ 
                           x_1+x_2+x_3+x_4+x_5=0\}
\end{equation}
\begin{equation} 
E''=\{ (x_1,x_2,x_3,x_4,x_5)\in \mathbb{E}_5\ |\ x_1=x_2=x_3=x_4=x_5\}
\end{equation}
\begin{equation}  
{\mathcal L}=(\pi +\pi ')(\mathbb{L})=\mathbb{Z}w_1+\mathbb{Z}w_2+\mathbb{Z}w_3+\mathbb{Z}w_4
\end{equation}
\begin{equation} \fl 
L=\mathbb{L}\cap {\mathcal E}=5{\mathcal L}=\{ (x_1,x_2,x_3,x_4,x_5)\in \mathbb{L}\ |\ 
                           x_1+x_2+x_3+x_4+x_5=0\}
\end{equation}
where
\begin{equation}\begin{array}{ll}
w_1=\frac{1}{\sqrt{10}} (4,-1,-1,-1,-1) \quad & 
w_2=\frac{1}{\sqrt{10}} (-1,4,-1,-1,-1) \\[2mm]
w_3=\frac{1}{\sqrt{10}} (-1,-1,4,-1,-1) \quad &
w_4=\frac{1}{\sqrt{10}} (-1,-1,-1,4,-1) .
\end{array}
\end{equation}

We can choose $z_j=(\kappa j,0,0,0,0)$ since
$\pi ''z_i\not=\pi ''z_j$ for $i\not=j$, 
\[  {\mathcal E}_j=z_j+{\mathcal E}=
\{ (x_1,x_2,x_3,x_4,x_5)\in \mathbb{E}_5\ |\ x_1+x_2+x_3+x_4+x_5=\kappa j\}\]
and
\[ \mathbb{L}\subset \bigcup_{j\in \mathbb{Z}}{\mathcal E}_j .\]
The set $\mathbb{K}\cap {\mathcal E}_i$ is non-empty only for $i\in \{ 0,1,2,3,4,5\}$,
but ${\mathcal K}_i=\pi '(\mathbb{K}\cap {\mathcal E}_i)$ has non-empty interior only for
$i\in \{ 1,2,3,4\}$. 
The set ${\mathcal K}_1$ is the regular pentagon with the vertices
\begin{equation}\begin{array}{l} 
\pi '(\kappa ,0,0,0,0)=\frac{1}{\sqrt{10}}\, (2,-\tau ,-\tau ',-\tau ', -\tau )\\[2mm]
\pi '(0,\kappa ,0,0,0)=\frac{1}{\sqrt{10}} (-\tau ,2,-\tau ,-\tau ',-\tau ')\\[2mm]
\pi '(0,0,\kappa ,0,0)=\frac{1}{\sqrt{10}} (-\tau ',-\tau ,2,-\tau ,-\tau ')\\[2mm]
\pi '(0,0,0,\kappa ,0)=\frac{1}{\sqrt{10}} (-\tau ',-\tau ',-\tau ,2,-\tau )\\[2mm]
\pi '(0,0,0,0,\kappa )=\frac{1}{\sqrt{10}} (-\tau ,-\tau ',-\tau ',-\tau ,2)
\end{array}
\end{equation}
${\mathcal K}_2=-\tau {\mathcal K}_1$, 
${\mathcal K}_3=\tau {\mathcal K}_1$,
${\mathcal K}_4=-{\mathcal K}_1$,  
and we can re-define the Penrose pattern ${\mathcal Q}$ 
as the Baake-Moody set
\begin{equation}
{\mathcal Q}=\bigcup_{i=1}^4
\left\{ \pi x\ \left| \ x\in {\mathcal L}_i , \ \pi ' x\in {\mathcal K}_i
\right.\right\}
\end{equation}
where ${\mathcal L}_j=(\pi +\pi ')z_j+L=jw_1+L.$ 
This definition is directly related to de Bruijn's definition \cite{M}.

\section{A 3D Penrose pattern}

The icosahedral group $Y=235=\left< a,\; b\ |\ a^5=b^2=(ab)^3=I\right>$
has five irreducible non-equivalent representations. Its character table is
\begin{equation}\label{table}
\begin{array}{ccccccc}
 & & 1\; e &  12\; a &  15\; b & 20\; ab & 12\; a^2\\
\Gamma _1 & & 1 & 1 & 1 & 1 & 1 \\
\Gamma _2 & & 3 & \tau & -1 & 0 & \tau ' \\
\Gamma _3 & & 3 & \tau ' & -1 & 0 & \tau \\
\Gamma _4 & & 4 & -1 & 0 & 1 & -1\\
\Gamma _5 & & 5 & 0 & 1 & -1 & 0.
\end{array} 
\end{equation}
A realization of $\Gamma _2$ in the Euclidean 3D space $\mathbb{E}_3$ is the
representation generated by the rotations $a,\ b :\mathbb{E}_3\longrightarrow \mathbb{E}_3$
\begin{equation}\label{Y} \fl \begin{array}{l}
a(\alpha ,\beta  ,\gamma )=
\left(\frac{\tau -1}{2}\alpha -\frac{\tau }{2}\beta  +\frac{1}{2}\gamma ,
\ \frac{\tau }{2}\alpha +\frac{1}{2}\beta  +\frac{\tau -1}{2}\gamma ,
\ -\frac{1}{2}\alpha +\frac{\tau -1}{2}\beta  
+\frac{\tau }{2}\gamma \right)\\[1mm]
b(\alpha ,\beta  ,\gamma )=(-\alpha ,-\beta  ,\gamma ).
\end{array} \end{equation}
If in relation (\ref{Y}) we replace $\tau $ by $\tau ' $ then we obtain a
realization of $\Gamma _3$.

Let $\varepsilon _1=(1,0,...,0),\ \varepsilon _2=(0,1,0,...,0),\ ...,\
\varepsilon _6=(0,...,0,1)$ be the canonical basis of $\mathbb{E}_6.$
The points of the one-shell $Y$-cluster  
\[ {\mathcal C}=Y(1,\tau ,0)=\{ e_1,e_2,...,e_6,-e_1,-e_2,...,-e_6\} \]
where 
\begin{equation}
\begin{array}{lll}
e_1=(1,\tau ,0) & e_3=(-\tau ,0,1) & e_5=(\tau ,0,1)\\
e_2=(-1,\tau ,0) & e_4=(0,-1,\tau ) & e_6=(0,1,\tau )
\end{array}
\end{equation}
are the vertices of a regular icosahedron. The action of $a$ and $b$ on 
the set ${\mathcal C}$ is described by the signed permutations
\begin{equation}
\fl  a=\left( \begin{array}{rrrrrr}
                e_1 & e_2 & e_3 & e_4 & e_5 & e_6\\
                e_2 & e_3 & e_4 & e_5 & e_1 & e_6 
                \end{array} \right)
\qquad b=\left( \begin{array}{rrrrrr}
                e_1 & e_2 & e_3 & e_4 & e_5 & e_6\\
                -e_1 & -e_2 & e_5 & e_6 & e_3 & e_4 
                \end{array} \right) 
\end{equation}
and the corresponding transformations $a,\ b:\mathbb{E}_6\longrightarrow \mathbb{E}_6$
\begin{equation}
\fl a=\left( \begin{array}{rrrrrr}
                \varepsilon _1 & \varepsilon _2 & \varepsilon _3 & 
                \varepsilon _4 & \varepsilon _5 & \varepsilon _6 \\
                \varepsilon _2 & \varepsilon _3 & \varepsilon _4 & 
                \varepsilon _5 & \varepsilon _1 & \varepsilon _6 
                \end{array} \right)
\qquad  b=\left( \begin{array}{rrrrrr}
     \varepsilon _1 & \varepsilon _2 & \varepsilon _3 & \varepsilon _4 & 
     \varepsilon _5 & \varepsilon _6\\
     -\varepsilon _1 & -\varepsilon _2 & \varepsilon _5 & \varepsilon _6 & 
     \varepsilon _3 & \varepsilon _4
                \end{array} \right) 
\end{equation}
generate the orthogonal representation of $Y$ in $\mathbb{E}_6$ 
\begin{equation} \begin{array}{l} 
a(x_1,x_2,x_3,x_4,x_5,x_6)=(x_5,x_1,x_2,x_3,x_4,x_6)\\ 
b(x_1,x_2,x_3,x_4,x_5,x_6)=(-x_1,-x_2,x_5,x_6,x_3,x_4). 
\end{array}  
\end{equation} 

The vectors 
\begin{equation}
\fl v_1=\varrho (1,-1,-\tau ,0,\tau ,0)\qquad
v_2=\varrho (\tau ,\tau ,0,-1,0,1)\qquad
v_3=\varrho (0,0,1,\tau ,1,\tau )
\end{equation}
where $\varrho =1/\sqrt{4+2\tau }$, form an orthonormal basis of 
the $Y$-invariant subspace
\begin{equation}
E=\{ (\langle r,e_1\rangle ,\langle r,e_2\rangle,...,
\langle r,e_6\rangle )\ |\ r\in \mathbb{E}_3\} .
\end{equation}
The isometry
\begin{equation}
\varphi :\mathbb{E}_3\longrightarrow E:\ r\mapsto 
 (\varrho \langle r,e_1\rangle ,\varrho \langle r,e_2\rangle,...,
\varrho \langle r,e_6\rangle )
\end{equation}
which is an isomorphism of representations \cite{C1,C2} of $Y$ 
with the property $\varphi (\alpha ,\beta ,\gamma )=
\alpha v_1+\beta v_2+\gamma v_3$ 
allows us to identify the `physical' space $\mathbb{E}_3$ with $E$.

The matrices  of the orthogonal projectors 
$\pi , \pi ^\perp :\mathbb{E}_6\longrightarrow \mathbb{E}_6$
corresponding to $E$ and 
$E^\perp =\{ x\in \mathbb{E}_6\ |\ \langle x,y\rangle =0\ {\rm for\ all\ }
y\in E \}$ in the basis 
$\{ \varepsilon _1,...,\varepsilon _6\}$ are 
\begin{equation}
\pi ={\mathcal M}(1/2,\sqrt{5}/10)\ \ \ \ \ \ \ \ 
\pi ^{\perp }={\mathcal M}(1/2,-\sqrt{5}/10)
\end{equation}
where
\begin{equation}
{\mathcal M}(\xi, \beta )=\left(
\begin{array}{rrrrrr}
\alpha &\beta &\beta &\beta &\beta &\beta \\
\beta &\alpha &\beta &-\beta &-\beta &\beta \\
\beta &\beta &\alpha &\beta &-\beta &-\beta \\
\beta &-\beta &\beta &\alpha &\beta &-\beta \\
\beta &-\beta &-\beta &\beta &\alpha &\beta \\
\beta &\beta &-\beta &-\beta &\beta &\alpha 
\end{array} \right).
\end{equation}
They can be obtained one from the other by using the 
transformation  $\sqrt{5}\mapsto -\sqrt{5}$.

Let $\kappa =1/\varrho $, $\mathbb{L}=\kappa \mathbb{Z}^6$, $\mathbb{K}=[0,\kappa ]^6$,
and let $K=\pi ^\perp (\mathbb{K})$. The pattern defined in terms of the 
strip projection method 
\begin{equation}
{\mathcal Q}=\left\{ \pi x\ |\ x\in {\mathcal L},\ \pi 'x\in K\right\}
\end{equation}
is the set of vertices of a {\em 3D Penrose (singular) pattern} \cite{K}. 
It can be regarded as a quasiperiodic packing of 
interpenetrating icosahedra. Only a very small part of the icosahedra
occurring in this pattern are fully occupied \cite{K}.

In this case, $E''=\{ 0\}$, ${\mathcal E}=\mathbb{E}_6$, ${\mathcal L}=\mathbb{L}$,
\begin{equation}\begin{array}{ccccccc}
\pi x \leftarrow x 
&:E& \stackrel{\pi }\longleftarrow & \mathbb{E}_6
& \stackrel{\pi ' }\longrightarrow & E' :& x \rightarrow \pi ' x\\
&&&\cup &&& \\
&&& \mathbb{L} &&&
\end{array}
\end{equation} 
is a cut and project scheme, and ${\mathcal Q}$ is a {\em model set} \cite{M}, 
that is a particular case of Baake-Moody set.

\section{A quasiperiodic packing of dodecahedra}

The points of the one-shell $Y$-cluster
\begin{equation}
{\mathcal C}=Y(1,1,1)=\{ e_1,e_2,...,e_{10},-e_1,-e_2,...,-e_{10}\}
\end{equation}
where 
\begin{equation}
\begin{array}{lll}
e_1=(1,1,1) & e_4=(1-\tau ,0,\tau ) & e_7=(\tau ,\tau -1,0)\\
e_2=(0,\tau ,\tau -1) & e_5=(\tau -1,0,\tau ) & e_8=(0,\tau ,1-\tau )\\
e_3=(-1,1,1) & e_6=(1,-1,1) & e_9=(-\tau ,\tau -1,0)\\
 & & e_{10}=(-1,-1,1)
\end{array}
\end{equation}
are the vertices of a regular dodecahedron. 
By using the method from the previous section and the 
canonical basis 
$\{ \varepsilon _1,\ \varepsilon _2,\ ...,\ \varepsilon _{10}\}$ 
of $\mathbb{E}_{10}$ we define the permutation representation
\begin{equation}\label{rep}
 \begin{array}{l} a=\left( \begin{array}{llllllllll}
\varepsilon _1 & \varepsilon _2 & \varepsilon _3 & \varepsilon _4 & 
\varepsilon _5 & \varepsilon _6 & \varepsilon _7 & \varepsilon _8 & 
\varepsilon _9 & \varepsilon _{10}\\[1mm]
\varepsilon _2 & \varepsilon _3 & \varepsilon _4 & \varepsilon _5 & 
\varepsilon _1 & \varepsilon _7 & \varepsilon _8 & \varepsilon _9 & 
\varepsilon _{10} & \varepsilon _6
                \end{array} \right)\\[5mm]
b=\left( \begin{array}{llllllllll}
\varepsilon _1 & \varepsilon _2 & \varepsilon _3 & \varepsilon _4 & 
\varepsilon _5 & \varepsilon _6 & \varepsilon _7 & \varepsilon _8 & 
\varepsilon _9 & \varepsilon _{10}\\[1mm]
\varepsilon _{10} & -\varepsilon _8 & \varepsilon _6 & \varepsilon _5 & 
\varepsilon _4 & \varepsilon _3 & -\varepsilon _7 & -\varepsilon _2 & 
-\varepsilon _9 & \varepsilon _1
                \end{array} \right) \end{array}
\end{equation}
of $Y$ in $\mathbb{E}_{10}$.

The vectors 
\begin{equation}\begin{array}{l}
v_1=\varrho (1,0,-1,1-\tau ,\tau -1,1,\tau ,0,-\tau ,-1)\\
v_2=\varrho (1,\tau ,1,0,0,-1,\tau -1,\tau ,\tau -1,-1)\\
v_3=\varrho (1,\tau -1,1,\tau ,\tau ,1,0,1-\tau ,0,1) \end{array}
\end{equation}
where $\varrho =1/\sqrt{10}$, form an orthonormal basis of the 
$Y$-invariant subspace
\begin{equation}
E=\{ (\langle r,e_1\rangle ,\langle r,e_2\rangle,...,
\langle r,e_{10}\rangle )\ |\ r\in \mathbb{E}_3\} 
\end{equation}
and the isometry
\begin{equation}
\varphi :\mathbb{E}_3\longrightarrow E:\ r\mapsto 
 (\varrho \langle r,e_1\rangle ,\varrho \langle r,e_2\rangle,...,
\varrho \langle r,e_{10}\rangle )
\end{equation}
which is an isomorphism of representations \cite{C2} of $Y$ 
with the property $\varphi (\alpha ,\beta ,\gamma )=
\alpha v_1+\beta v_2+\gamma v_3$ 
allows us to identify the `physical' space $\mathbb{E}_3$ with $E$.

The projectors corresponding to $E$ and 
$E^\perp =\{ x\in \mathbb{E}_{10}\ |\ \langle x,y\rangle =0\ {\rm for\ all\ }
y\in E \}$ are
\begin{equation}
\pi ={\mathcal M}\left(\frac{3}{10},\frac{\sqrt{5}}{10},\frac{1}{10}\right)\qquad
\pi ^\perp ={\mathcal M}\left(\frac{7}{10},-\frac{\sqrt{5}}{10},-\frac{1}{10}\right)\
\end{equation}
where
\begin{equation}\fl 
{\mathcal M}(\alpha ,\beta , \gamma )=\left(
\begin{array}{rrrrrrrrrr}
\alpha & \beta & \gamma & \gamma & \beta & \gamma & \beta & \gamma 
&-\gamma &-\gamma \\
\beta & \alpha & \beta & \gamma & \gamma &-\gamma & \gamma & \beta 
& \gamma &-\gamma \\
\gamma & \beta & \alpha & \beta & \gamma &-\gamma &-\gamma & \gamma 
& \beta & \gamma \\
\gamma & \gamma & \beta & \alpha & \beta & \gamma &-\gamma &-\gamma 
& \gamma & \beta \\
\beta & \gamma & \gamma & \beta & \alpha & \beta & \gamma &-\gamma 
&-\gamma & \gamma \\
\gamma &-\gamma &-\gamma & \gamma & \beta & \alpha & \gamma &-\beta 
&-\beta & \gamma \\
\beta & \gamma &-\gamma &-\gamma & \gamma & \gamma & \alpha & \gamma 
&-\beta &-\beta \\
\gamma & \beta & \gamma &-\gamma &-\gamma &-\beta & \gamma & \alpha 
& \gamma &-\beta \\
-\gamma & \gamma & \beta & \gamma &-\gamma &-\beta &-\beta & \gamma 
& \alpha & \gamma \\
-\gamma &-\gamma & \gamma & \beta & \gamma & \gamma &-\beta &-\beta 
& \gamma & \alpha 
\end{array} \right) .
\end{equation}

Let $\kappa =1/\varrho $, $\mathbb{L}=\kappa \mathbb{Z}^{10}$, $\mathbb{K}=[0,\kappa ]^{10}$,
and $K=\pi ^\perp (\mathbb{K})$.
Following the analogy with the Penrose case we define the icosahedral 
pattern
\begin{equation}\label{pattern}
{\mathcal Q}=\left\{ \pi x\ |\ x\in \mathbb{L},\ \pi ^\perp x\in K\right\}.
\end{equation}
Since the arithmetic neighbours of a point ${\pi x\in \mathcal Q}$ are 
distributed on the vertices of the regular dodecahedron 
$\pi x+{\mathcal C}$, the pattern ${\mathcal Q}$ can be regarded as a
quasiperiodic packing of interpenetrating dodecahedra. 
Evidently, only a very small part of the dodecahedra occurring in this pattern
can be fully occupied. 

The pattern ${\mathcal Q}$ can be re-defined as a Baake-Moody set by using 
the decomposition $E^\perp =E'\oplus E''$, where $E'$ and $E''$
are the subspaces corresponding to the orthogonal projectors
\begin{equation}
\pi '={\mathcal M}\left(\frac{3}{10},-\frac{\sqrt{5}}{10},\frac{1}{10}\right)\qquad
\pi ''={\mathcal M}\left(\frac{2}{5},0,-\frac{1}{5}\right) .
\end{equation}
The projectors $\pi $ and $\pi '$ can be obtained one from the other by
using the transformation $\sqrt{5}\mapsto -\sqrt{5}$.
The superspace ${\mathcal E}=E\oplus E'$ is six-dimensional, but in order to 
use it we have to determin the `atomic surfaces' ${\mathcal K}_i$.

\section{Concluding remarks}

If in the previous example we replace the starting cluster $Y(1,1,1)$ by 
the icosidodecahedron $Y(1,0,0)$ then we get a quasiperiodic packing of
icosidodecahedra. If we replace it by the two-shell $Y$-cluster
\[
\fl {\mathcal C}=Y\{ \alpha (1,\tau ,0),\, \beta (1,1,1)\}=
Y(\alpha ,\alpha \tau ,0)\, \cup \, Y(\beta ,\beta ,\beta )=
\{ e_1,...,e_{16},-e_1,...,-e_{16}\}
\]
where $\alpha $, $\beta $ are rational positive numbers, then we get a pattern 
${\mathcal Q}$ such that 
the arithmetic neighbours of each point $\pi x\in {\mathcal Q}$
are distributed \cite{C2} on two shells, 
namely, on the vertices of a regular 
icosahedron of radius $\alpha \sqrt{\tau +2}$ and on the vertices of
a regular dodecahedron of radius $\beta \sqrt{3}$. 
The  pattern ${\mathcal Q}$ can be regarded
as a quasiperiodic packing of interpenetrating copies of ${\mathcal C}$. 
Only a very small part of the copies of the cluster 
${\mathcal C}$ occurring in ${\mathcal Q}$ can be fully occupied. 
We think that the frequency of occurrence of the fully
occupied icosahedra may be much greater in this pattern than in the 3D
Penrose pattern.
The pattern ${\mathcal Q}$ can be re-defined as a Baake-Moody set by using
a 6D superspace, but we have to determin some rather complicated `atomic surfaces'
${\mathcal K}_i$. 

From our general theory it follows that the permutation representation 
defined in $\mathbb{E}_k$ by any $Y$-cluster ${\mathcal C}$ contains the 
irreducible representation $\Gamma _2$, and the corresponding $Y$-invariant
subspace $E$ can be determined explicitly. If we change the sign of
$\sqrt{5}$ in the expression of the orthogonal projector $\pi $ corresponding
to $E$  we get the orthogonal projector $\pi '$ corresponding to another 3D
$Y$-invariant subspace $E'$. The subduced representation of $Y$ in $E'$ 
belongs to $\Gamma _3$, and the orthogonal projector
$\pi +\pi '$ corresponding to ${\mathcal E}=E\oplus E'$ has rational entries.
A quasiperiodic packing ${\mathcal Q}$ of copies of ${\mathcal C}$ can be
defined in a natural way in terms of  the strip projection method by using
the decomposition $\mathbb{E}_k=E\oplus E^\perp $. The same pattern can be 
re-defined as a Baake-Moody set in the 6D  superspace 
${\mathcal E}=E\oplus E'$ by using some rather complicated `atomic surfaces'
${\mathcal K}_i$.

\end{document}